\documentclass[letter,tradiabstract]{aa} 
\usepackage{graphicx}
\usepackage{txfonts}
\begin{document}
\title{Discovery of Eclipsing Binary Central Stars in the Planetary Nebulae M 3-16, H 2-29 and M 2-19}
\titlerunning{Eclipsing Binary Central Stars of M 3-16, H 2-29 and M 2-19}
   \author{B. Miszalski
           \inst{1,2}
           \and
           A. Acker
           \inst{1}
           \and
           A.F.J. Moffat
           \inst{3}
           \and
           Q. A. Parker
           \inst{2,4}
           \and
           A. Udalski
           \inst{5}
          }
\institute{Observatoire Astronomique, Universit\'e Louis Pasteur, 67000, Strasbourg, France\\
\email{brent@newb6.u-strasbg.fr, acker@newb6.u-strasbg.fr}
         \and
Department of Physics, Macquarie University, Sydney, NSW 2109, Australia\\ 
             \email{qap@ics.mq.edu.au}
         \and
         D\'ept. de physique, Univ. de Montr\'eal C.P. 6128, Succ. Centre-Ville, Montr\'eal, QC H3C 3J7, Canada\\
         \email{moffat@astro.umontreal.ca}
         \and
         Anglo-Australian Observatory, Epping, NSW 1710, Australia
         \and
         Warsaw University Observatory, Al. Ujazdowskie 4, PL-00-478, Warsaw, Poland\\
         \email{udalski@astrouw.edu.pl}
         }
   \date{Received -; accepted -}
   \abstract{
   Progress in understanding the formation and evolution of planetary nebulae (PN) has been restricted by a paucity of well-determined central star masses. 
   To address this deficiency we aim to (i) significantly increase the number of known eclipsing binary central stars of PN (CSPN), and subsequently (ii) directly obtain their masses and absolute dimensions by combining their light-curve parameters with planned radial velocity data.
   Using photometric data from the third phase of the Optical Gravitational Lensing Experiment (OGLE) we have searched for periodic variability in a large sample of PN towards the Galactic Bulge using Fourier and phase-dispersion minimisation techniques. 
   Among some dozen periodically variable CSPN found, we report here on three new eclipsing binaries: M 3-16, H 2-29 and M 2-19. 
   We present images, confirmatory spectroscopy and light-curves of the systems.
   }
   \keywords{ binaries: eclipsing - planetary nebulae: general
               }
   \maketitle
\section{Introduction}
The key parameter for any star is its mass. The least model-dependent method to obtain masses is via Keplerian orbits in binary systems. Eclipsing binaries provide critical information necessary to secure the masses and other fundamental stellar parameters. 
The mass is obtained from radial velocity (RV) orbits of both stars combined with the orbital inclination from photometric eclipses.

Yet directly derived masses have not been fully exploited in the case of central stars of planetary nebulae (CSPN), partly because so few eclipsing binary systems are known. Indirect methods, such as nebula modelling (e.g. Gesicki et al. 2006) or model atmosphere fitting (e.g. Napiwotzki 1999), offer a poor alternative to relatively model-independent techniques such as Keplerian orbits. 

Obtaining direct masses for a sufficiently large number of CSPN is key to understanding the link between the AGB phase of stellar evolution and the final white dwarf stage. Such a study will also greatly improve our understanding of PN shaping mechanisms (e.g. Zijlstra 2007) and can provide compelling candidates for SN Ia progenitors (Tovmassian et al. 2008). Because most stars of low-intermediate mass experience the PN stage, PN are also major contributors to ISM enrichment and the chemical evolution of galaxies.

De Marco et al. (2008, and references therein) summarised the status of photometrically variable CSPN, including 12 close binaries with known periods. 
Three of these are eclipsing: UU Sge (Pollaco \& Bell 1993), V477 Lyr (Pollaco \& Bell 1994), and BE UMa (Ferguson et al. 1999).
All three have viable mass determinations using both RV and light-curves, with 0.5--0.7 M$_\odot$ for the hot 60--120 kK primary and 0.15--0.36 M$_\odot$ for the irradiated 5--7 kK secondary. 
Their light-curves display strong irradiation effects, in addition to typical emission lines from the heated hemisphere of the secondary.
All these systems are believed to be post-CE (common-envelope) binaries, where the secondary spirals in through the primary's AGB envelope.
The remaining stars of De Marco et al. 2008 are non-eclipsing and therefore any mass estimates for them are subject to considerable uncertainty.

Either RV or photometric monitoring surveys may be used to \emph{find} CSPN binaries. 
However, if the ultimate goal (as in this work) is to get the masses from binaries, then it is easier and more efficient to start with photometric surveys to preselect eclipsing binaries, for which RV orbits combined with photometrically-derived inclinations will secure the masses. 
Indeed, RV monitoring surveys have proven to be difficult with many candidates showing RV variability without definitive periodicity (e.g. De Marco et al. 2004). 

This paper introduces a novel use of extant online photometric data from the OGLE microlensing survey. We describe in Section 2 a search for periodic variability in over 300 PN towards the Galactic Bulge. In Section 3 we present the discovery of three new eclipsing binary CSPN. We discuss the nature of these new discoveries in Section 4 and conclude in Section 5.

\section{CSPN Variability towards the Galactic Bulge}
\subsection{OGLE-III}
To address the paucity of binary CSPN we have searched for periodic photometric variations in data from the Optical Gravitational Lensing Experiment (OGLE; Udalski et al. 1992) in its most recent third phase (OGLE-III; e.g. Udalski et al. 2002). 
Each phase of the OGLE project improves its temporal and areal coverage of the Galactic Bulge and Magellanic Clouds over previous phases. 
The areal coverage of OGLE-III is very well-matched to the PN distribution towards the Bulge (e.g. the most frequently sampled fields cover $|\ell|\le5$ and $-5 < b < -2$).

The OGLE-III survey uses the 1.3-m Warsaw telescope at the Las Campanas Observatory, Chile. It is equipped with a CCD mosaic camera with eight 2K x 4K CCDs giving a 35 $\times$ 35.5 arcmin$^2$ field of view with 0.26\arcsec\, pixels.
The survey is conducted in the I-band reaching a limiting magnitude of I$\sim$20, though some V-band reference images are also available. 
The data span approximately six years starting from June 2001 with gaps between each Bulge season.

\subsection{Search Method}
Surveys for PN towards the Galactic Bulge have recently been considerably enhanced by the Macquarie/AAO/Strasbourg H$\alpha$ PN catalogues (MASH; Parker et al. 2006, Miszalski et al. 2008). 
By effectively doubling the PN population towards the Bulge, MASH now enables a large population of Galactic PN to be monitored for variability in a homogeneous fashion by one photometric survey.

For each of the over 300 PN that fell in the OGLE-III survey fields, we examined OGLE-III I-band finder charts to identify central stars for which to extract time-series photometry. 
Approximately 30\% of these had CSPN identified unambiguously with the help of significantly lower nebular contamination in the I-band.
In more difficult cases we selected a number of likely candidates close to the geometric centre of each PN.

Time series photometry for each identified CSPN was subsequently retrieved from the OGLE-III database generated by a data reduction pipeline based on the difference image analysis (DIA) method (Wo\'zniak 2000). 
The stellar-optimised pipeline could not extract some light-curves in cases of nebular contamination.
We examined the photometric data using both the Fourier analysis package \textsc{PERIOD04} (Lenz \& Breger 2004) and the phase dispersion minimisation (PDM) IRAF task (Stellingwerf 1978). 
Initially the data were searched with the Fourier technique for frequencies in the range 0--50 d$^{-1}$ (the latter being the typical Nyquist frequency for well-sampled fields) with a step rate 2.2 $\times$ 10$^{-4}$ d$^{-1}$ ($\sim$1/5 times the inverse sample length of $\sim$6 years). 
Periodicity was identified when the Fourier peak amplitude reached at least 3$\sigma$ above the noise. 
The final selection of the best frequency was made after visual inspection of the phased light-curve. We used the PDM technique to check our Fourier results for $P$$\sim$0.01--4 days.

\section{Results}
Some dozen periodically variable CSPN were found during our examination of OGLE-III data. 
A photometric binary fraction of 10--15\% (Bond 2000) compares reasonably well with our results, 
but complex selection criteria involving CSPN identification and uneven field sampling precludes a meaningful estimate of our binary fraction from being given here. A future paper will be dedicated to this topic (Miszalski et al., in prep).

\subsection{Three New Eclipsing Binary CSPN}
\begin{figure*}
   \begin{center}

      \includegraphics[scale=1.0]{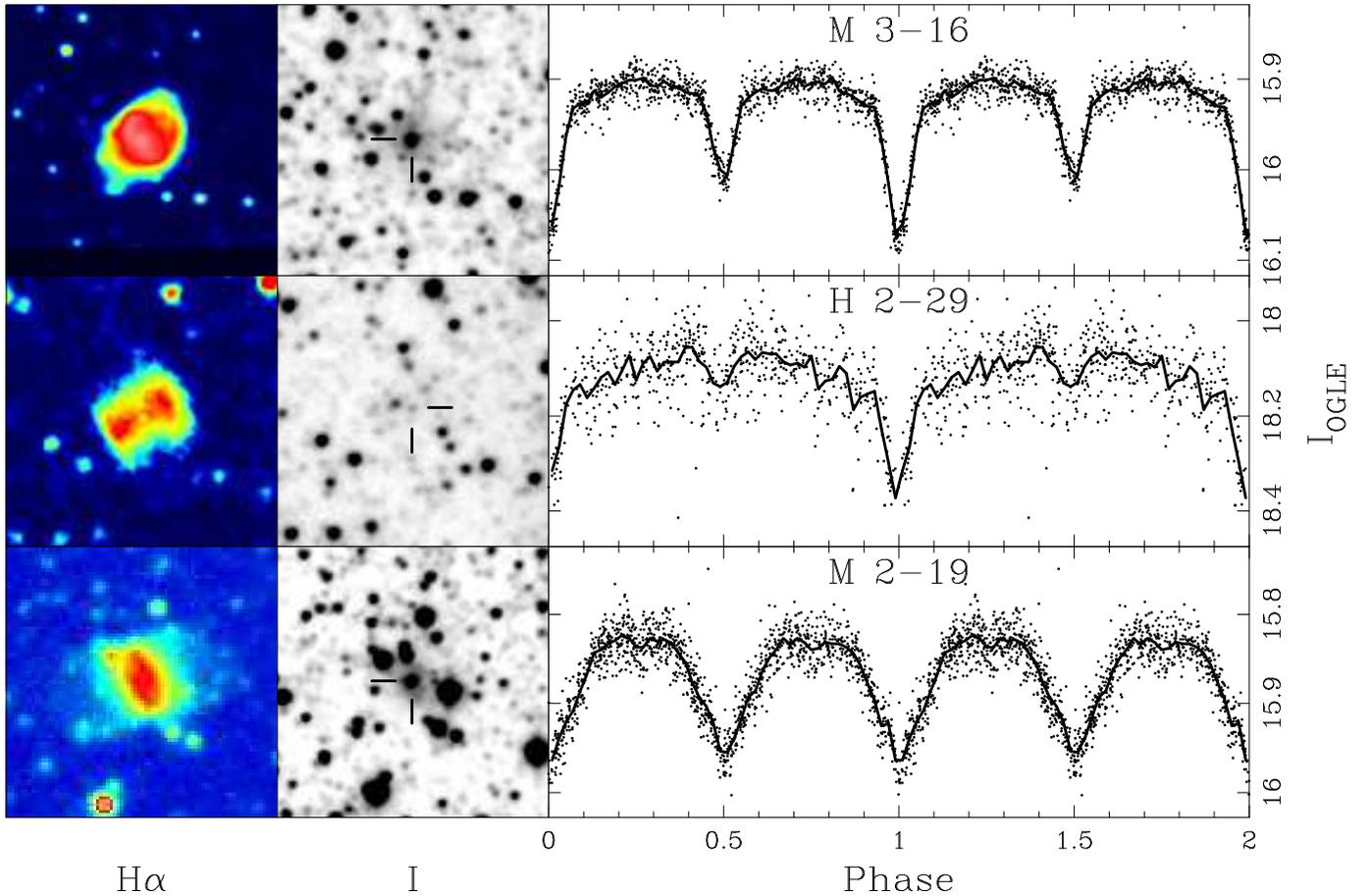}
   \end{center}
   \caption{A montage of the three newly discovered eclipsing systems. 
   Each row depicts from left to right: An H$\alpha$ image of the nebula, an OGLE-III I-band image, and a phased OGLE-III light-curve of the central star (marked on the I-band image). Periods and ephemerides are given in Table \ref{tab:basic}.
   All images are 30 $\times$ 30 arcsec$^2$ with North to top and East to left. 
   The solid curve is a binned light-curve representation sampled at $\delta\phi=0.01$. 
   }
   \label{fig:lc}
\end{figure*}

Of all the periodically variable CSPN found, those of M 3-16 (PN G359.1-02.3), H 2-29 (PN G357.6-03.3) and M 2-19 (PN G000.2-01.9) stand out as having clear eclipsing light-curves. 
These PN have been known for some time (Acker et al. 1992) but are all relatively unstudied (though see Section \ref{sec:spec}).
The other variables, including some MASH CSPN, tend to exhibit sinusoidal-like light-curves and will be discussed elsewhere.
In Figure \ref{fig:lc} we present for each object an H$\alpha$ image, the OGLE-III I-band image and its OGLE-III light-curve. The data are described in the following sections.

\subsection{Photometry}
\subsubsection{H$\alpha$ Images}
High-quality H$\alpha$ images of each nebula are presented in the first column of Figure \ref{fig:lc}. 
We obtained an H$\alpha$ image of M 3-16 using the Mosaic II camera on the CTIO 4-m Blanco telescope in June 2008. 
The H 2-29 and M 2-19 images were taken with ESO NTT/EMMI during programs 67.D-0527(A) and 71.D-0448(A) (Ruffle et al. 2004). 
Note that all images include both adjacent [NII] emission lines in the bandpass of the H$\alpha$ filters. In all instances the resolution is well-matched to the OGLE-III resolution of 0.26\arcsec/pixel.

\subsubsection{CSPN Identification}
Identification of central stars in the OGLE-III I-band images requires careful scrutiny and a full discussion of selection effects will be given in Miszalski et al. (in prep).
Nevertheless, in each of our three eclipsing binary CSPN we have unambiguously identified the CSPN, as indicated by the markers in Figure \ref{fig:lc}. 
In the case of M 3-16 and M 2-19 we see faint nebulae on the I-band images with centrally located stars. 
Although no I-band nebular emission can be seen for H 2-29, the star marked is centrally located with respect to the nebula in the H$\alpha$ image. Furthermore, we note that the CSPN marked is brighter in the OGLE-III V- than the I-band image, strongly suggesting that the star is blue and is therefore the actual CSPN.

\subsubsection{Light Curves}
Phased light-curves of each CSPN are given in Figure \ref{fig:lc}, with periods and ephemerides given in Table \ref{tab:basic}. 
No significant change in period was detected between each individual Bulge season for each object.

We visually inspected the light-curves phased with the period recovered by the Fourier method.
We doubled the period of M 3-16 to separate the overlapping, unequal eclipses. 
The slightly rounded shape of its light-curve between minima is interpreted as ellipsoidal variations.
The light-curve of H 2-29 phased with the initial Fourier period shows a moderately weak secondary minimum at 0.5 phase leading us to accept it as the most likely period. However, if the secondary minimum turns out to be spurious, then the true period would likely be twice our value.
When phased with the Fourier period, the light-curve of M 2-19 lacks a convincing secondary minimum at 0.5 phase and shows an unusually wide primary minimum. Therefore, we doubled the Fourier period of M 2-19.

\subsection{Spectroscopy}
\label{sec:spec}
Although nebular spectroscopy exists for all our objects, no identification of CSPN features have been reported in the literature (Cuisinier et al. 2000; Exter et al. 2004).
Spectroscopic observations of M 3-16 and M 2-19 were made with the AAOmega system on the 3.9-m Anglo-Australian Telescope (AAT), on the nights of 16 March 2008 and 29 May 2008. AAOmega is an optical multi-object spectrograph (Sharp et al. 2006) that is fibre-fed by the 2dF fibre positioner (Lewis et al. 2002). 
The data were taken in service mode and exposure times were 2 $\times$ 1500s for both M 3-16 and M 2-19, with an additional 1800s on M 2-19. 
We used the 580V and 385R volume-phase holographic (VPH) gratings for the blue and red arms, respectively. 
This gave a central resolution of 3.5 \AA\, in the blue and 5.3 \AA\, in the red and a wavelength coverage of 3700--5800 \AA\, and 5720--8850 \AA, respectively.

\begin{figure}
   \begin{center}

      \includegraphics[angle=270,scale=0.35]{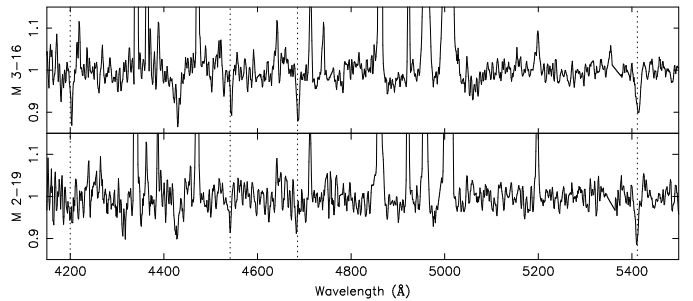}
   \end{center}
   \caption{AAOmega spectroscopy for M 3-16 (top) and M 2-19 (bottom) showing stellar HeII absorption lines (dotted lines). Contaminating nebular emission lines are also present. The spectra have been rectified and smoothed. 
   } 
   \label{fig:lines}
\end{figure}

The two epochs for M 3-16 and three epochs for M 2-19 were summed to give a S/N in the continuum at 5300 \AA\, of $\sim$40 and $\sim$30, respectively.
Both spectra exhibit stellar HeII absorption lines expected for hot O-type central stars (Figure \ref{fig:lines}). 
Evidence for stellar lines that betray a cool secondary component is weak. 
The CaII triplet feature is not seen and the possible presence of very faint N III and C III emission lines at $\sim$4640 \AA\, requires confirmation (see Pollacco \& Bell 1993, 1994). These features are more likely to be present in H 2-29, but no sufficiently deep spectra currently exist.
Our \emph{confirmatory} spectra of low S/N (individually) and low resolution were not intended to uncover radial velocity variations. 
Nevertheless, despite large uncertainties, we noticed no variation in the radial velocity of the HeII lines with respect to nebular lines. 
Higher resolution data is needed to place limits on any possible variation.

\begin{table*}
   \centering
   \caption{Basic data for the three new eclipsing binaries. 
   An average I-band magnitude $\bar{m}_I$ is calculated from the OGLE-III time-series photometry of $N$ observations over six Galactic Bulge seasons. The epoch of primary minimum $E_0$ is also given. The current zeropoint of $\bar{m}_I$ is roughly calibrated to an uncertainty of $\sim$0.1--0.2 mag.
   }
   \begin{tabular}{lllllllll}
      PN G & Name & RA (J2000) & Dec (J2000) & OGLE Field  & $E_0$ (HJD-2450000) & $P$ (days) & $\bar{m}_I$ & $N$ \\
      \hline
      359.1-02.3 & M 3-16 & 17 52 46.1 & -30 49 35 & BLG180.1  &3850.9184 $\pm$ 0.0015 & 0.573648 $\pm$ 0.000020 & 15.93 & 1039\\
      357.6-03.3 & H 2-29 & 17 53 16.8 & -32 40 38 & BLG155.1  &3619.5208 $\pm$ 0.0025 & 0.244110 $\pm$ 0.000005 & 18.12 & 609\\
      000.2-01.9 & M 2-19 & 17 53 45.6 & -29 43 46 & BLG101.3  &4224.7822 $\pm$ 0.0035 & 0.670170 $\pm$ 0.000020 & 15.87 & 1273 \\
      \hline
   \end{tabular}
   \label{tab:basic}
\end{table*}

\section{Discussion}
\subsection{Nature of the Systems}

At this stage we can only comment on the components of each of our new systems and leave analysis using the Wilson-Devinney method (Wilson \& Devinney 1971) to future papers. 
The periods of our systems are short and are within the range of the periods known for the majority of binary CSPN (De Marco et al. 2008).
Our light-curves show none of the very strong irradiation effects seen for UU Sge, V477 Lyr or BE UMa, suggesting a very different system composition, though irradiation effects may be present in H 2-29, albeit weaker and uncertain.
For M 2-19 and M 3-16 in particular, the eclipses are of similar and modest depths suggesting comparable effective temperatures of both components. 

Our low-resolution confirmatory spectroscopy show no detectable difference between radial velocities of HeII absorption lines and nebular lines in M 2-19 and M 3-16.
This is most likely due to our very limited sensitivity to radial velocity variations of up to 50--150 km s$^{-1}$ that are seen in known eclipsing CSPN (Pollacco \& Bell 1993, 1994). 
If such large variations are not confirmed by future monitoring, then we may be seeing the hot, ionising component of a triple system, with eclipses generated by two main-sequence stars of similar effective temperatures. 
This scenario is supported by the lack of strong irradiation effects and the similar eclipse depth in the light-curves of M 2-19 and M 3-16. However, we do not see any convincing stellar features indicative of this scenario in our low S/N spectra.
Such a case has been suggested for SuWt~2 that has a similar light-curve to M 3-16, but its bright A-type double-lined spectroscopic and eclipsing binary CSPN has not yet been ruled out as a foreground object (Bond et al. 2000).
There may be another explanation and more radial velocity data at higher resolution are needed to resolve the issue.

\subsection{Morphology}
The morphologies of our three nebulae all show bipolar features. 
This is consistent with the tendency for close binary stars to eject non-spherical nebulae through interacting stellar winds and tidal effects (e.g. Soker 1997). Indeed, the morphology of H 2-29 has been considered as a likely product of CE evolution with a stellar companion (Soker 1997).

The `butterfly' shape has been proposed to be the preferred morphology of nebulae ejected by close binaries because it is the most efficient transfer mechanism of angular momentum to the ejecta (Zijlstra 2007). 
M 2-19 exhibits such a classical `butterfly' morphology which is the first firmly established case with a considerably short period supporting this hypothesis. 
Other `butterfly' nebulae with a binary nucleus include NGC 2346 (Mendez et al. 1982), but it has a longer period of 16 days.

M 3-16 exhibits the most complex morphology of the three.
Surrounding the south-eastern side of the central star we see what may be part of an equatorial torus-like structure.
Perpendicular to this we have two arms often seen in other bipolar nebulae extended towards the north-west. If we use this to define an axis of symmetry, then 25 degrees away from this axis we have two opposing broad jets or ansae. 

\section{Conclusions}
We have doubled the known population of eclipsing binary CSPN after searching OGLE-III photometry in a large PN catalogue towards the Galactic Bulge.
The post-CE CSPN show bipolar morphologies consistent with the current hypothesis that close binaries lead to non-spherical nebulae. 
A following paper will present more new close binary CSPN, discuss the selection effects of the search in detail and provide a new, independent estimate the binary fraction of PN. 
Planned high-resolution spectroscopy will refine system parameters and derive masses of our new eclipsing binaries.

\begin{acknowledgements}
   BM acknowledges the support of an Australian Postgraduate Award and support from PICS/ULP.
   Thanks to Paul Dobbie for taking some of the AAOmega service observations and to David Frew for valuable discussions.
   AFJM is grateful for financial assistance to NSERC (Canada) and FQRNT (Qu\'ebec).
   The OGLE project is partially supported by the Polish MNiSW grant N20303032/4275.
   We thank the referee for very helpful comments.
\end{acknowledgements}


\begin{thebibliography}{}
  \bibitem[A92]{A92} Acker, A., Marcout, J., Ochsenbein, F., Stenholm, B., \& Tylenda, R.\ 1992, Garching: European Southern Observatory
  \bibitem[B00]{B00} Bond, H.~E.\ 2000, Asymmetrical Planetary Nebulae II: From Origins to Microstructures, 199, 115 
  \bibitem[C00]{C00} Cuisinier, F., Maciel, W.~J., K{\"o}ppen, J., Acker, A., \& Stenholm, B.\ 2000, A\&A, 353, 543
  \bibitem[DM04]{DM04} De Marco, O., Bond, H.~E., Harmer, D., et al. 2004, ApJL, 602, L93 
  \bibitem[DM]{dm08} De Marco, O., Hillwig, T.~C., \& Smith, A.~J.\ 2008, AJ, 136,323
  \bibitem[Ex04)]{E04} Exter, K.~M., Barlow, M.~J., \& Walton, N.~A.\ 2004, MNRAS, 349, 1291 
  \bibitem[F99]{F99} Ferguson, D.~H., Liebert, J., Haas, S., et al.\ 1999, ApJ, 518, 866 
  \bibitem[Gesicki et al.(2006)]{g06} Gesicki, K., Zijlstra, A.~A., Acker, A., et al.\ 2006, A\&A, 451, 925 
 \bibitem[LB04]{LB04} Lenz, P., \& Breger, M.\ 2004, The A-Star Puzzle, 224, 786 
 \bibitem[L02]{L02} Lewis, I.~J., Cannon, R.~D., Taylor, K., et al.\ 2002, MNRAS, 333, 279 
 \bibitem[1982]{1982M}Mendez, R.~H., Gathier, R., \& Niemela, V.~S.\ 1982, A\&A, 116, L5
  \bibitem[2008]{m08} Miszalski, B., Parker, Q.~A., Acker, A., et al.\ 2008, MNRAS, 384, 525
  \bibitem[Napiwotzki(1999)]{n99} Napiwotzki, R.\ 1999, A\&A, 350, 101 
  \bibitem[2008]{p06} Parker, Q.~A., Acker, A., Frew, D.~J., et al.\ 2006, MNRAS, 373, 79
  \bibitem[PB93]{pb93} Pollacco, D.~L., \& Bell, S.~A.\ 1993, MNRAS, 262, 377 
  \bibitem[PB94]{pb94} Pollacco, D.~L., \& Bell, S.~A.\ 1994, MNRAS, 267, 452 
  \bibitem[R04]{R04} Ruffle, P.~M.~E., Zijlstra, A.~A., Walsh, J.~R., et al.\ 2004, MNRAS, 353, 796 
  \bibitem[S06]{S06} Sharp, R., Saunders, W., Smith, G., et al. 2006, SPIE, 6269, 14
  \bibitem[S97]{S97} Soker, N.\ 1997, ApJS, 112, 487 
  \bibitem[S78]{S78} Stellingwerf, R.~F.\ 1978, ApJ, 224, 953 
  \bibitem[Tovmassian et al.(2008)]{t08} Tovmassian, G., Tomsick, J., Napiwotzki, R., et al.\ 2008, Astrophysics of Compact Objects, 968, 62 

  \bibitem[U92]{u92} Udalski, A., Szymanski, M., Kaluzny, J., et al. 1992, Acta Astronomica, 42, 253 
  \bibitem[2002]{u02} Udalski, A., Paczynski, B., Zebrun, K., et al. 2002, Acta Astronomica, 52, 217
  \bibitem[WD71]{WD71} Wilson, R.~E., \& Devinney, E.~J., 1971, ApJ, 166, 605
  \bibitem[W00]{W00} Wozniak, P.~R.\ 2000, Acta Astronomica, 50, 421 
 \bibitem[Zijlstra(2007)]{z07} Zijlstra, A.~A.\ 2007, Baltic Astronomy, 16, 79 
\end{thebibliography}
\end{document}